%% file: main.tex
\documentclass{article}
\usepackage{spconf,amsmath,graphicx}
\usepackage{float}
\usepackage{array}
\usepackage{multirow}
\usepackage{adjustbox}
\usepackage{graphicx}
\usepackage{tabularx}
\usepackage{makecell}
\usepackage{adjustbox}
\usepackage{array}
\usepackage{amsfonts}
\usepackage{tabularx}
\usepackage{subcaption}
\usepackage{enumitem}

\usepackage{multirow}
\usepackage{booktabs}
\usepackage{enumitem}
\setlist{nosep, leftmargin=14pt}

\usepackage{mwe} % to get dummy images

\usepackage{lipsum}

\usepackage[table]{xcolor}
\usepackage{colortbl}
\usepackage{graphicx}

\definecolor{pastelblue}{RGB}{192,228,236}   % Softer blue
\definecolor{pastelorange}{RGB}{255,223,186}  % Subtle orange instead of yellow
\definecolor{pastelgreen}{RGB}{188,238,188}   % Softer green
\definecolor{pastelviolet}{HTML}{CCC0DA}

\usepackage{tikz}

\usepackage{soul}

\title{Synthetic Volumetric Data Generation Enables Zero-Shot Generalization of Foundation Models in 3D Medical Image Segmentation}
\name{  Satrajit Chakrabarty$^{1*}$, Sourya Sengupta$^{1,2*\dagger}$, Gopal Avinash$^{1}$, Ravi Soni$^{1}$
\thanks{$^*$ Equally contributing first authors}
\thanks{$^\dagger$ This work was done during the author's internship at GE HealthCare.}
}

\address{$^{1}$GE HealthCare, San Ramon, CA, USA \\
$^{2}$University of Illinois Urbana–Champaign, Urbana,
IL, USA \\
}

\usepackage{fancyhdr}

\fancypagestyle{firstpage}{
    \fancyhf{} % Clear all headers and footers
    \chead{\small \it Paper accepted to IEEE ISBI 2026}
}

\begin{document}
\maketitle
\thispagestyle{firstpage} % <--- This line forces the header onto the first page

\begin{abstract}
Foundation models such as Segment Anything Model 2 (SAM 2) exhibit strong generalization on natural images and videos but perform poorly on medical data due to differences in appearance statistics, imaging physics, and three-dimensional structure. To address this gap, we introduce SynthFM-3D, an analytical framework that mathematically models 3D variability in anatomy, contrast, boundary definition, and noise to generate synthetic data for training promptable segmentation models without real annotations. We fine-tuned SAM 2 on 10,000 SynthFM-3D volumes and evaluated it on eleven anatomical structures across three medical imaging modalities (CT, MR, ultrasound) from five public datasets. SynthFM-3D training led to consistent and statistically significant Dice score improvements over the pretrained SAM 2 baseline, demonstrating stronger zero-shot generalization across modalities. When compared with the supervised SAM-Med3D model on unseen cardiac ultrasound data, SynthFM-3D achieved 2–3× higher Dice scores, establishing analytical 3D data modeling as an effective pathway to modality-agnostic medical segmentation.
\end{abstract}
\begin{keywords}
Synthetic data, Interactive segmentation, Foundation models, Zero-shot, Segment Anything Model
\end{keywords}

% \begin{figure*}[!t]
%     \centering
%     % \fbox{\rule{0pt}{5in}\rule{\linewidth}{0pt}}
%     \includegraphics[width=0.7\textwidth]{figures/train.png} 
%     \caption{Overview of the SynthFM-3D data generation process. The top row illustrates the method, where a 2D mask is volumetrically extrapolated into 3D label maps and corresponding synthetic image volumes through morphological and appearance transformations. The bottom row shows the resulting diversity—multiple texture and contrast variations generated from the same underlying mask—demonstrating the controllable variability in SynthFM-3D’s training data.
% }
% \end{figure*}

%%%%%%%%%%%%%%%%%%%%%%%%%%%%%%%%%%%%%%%%%%%%%%%%%%%%%%%%%%%
\section{Introduction}\label{sec:intro}
Foundation models (FMs) have recently transformed image and video segmentation, achieving state-of-the-art performance across interactive tasks. Meta’s Segment Anything Model 2 (SAM~2)~\cite{ravi2024sam}, a promptable video object segmentation framework built on memory-based transformers, generalizes well to natural images and videos but performs suboptimally on medical data due to domain differences in texture, contrast, and imaging physics.

Accurate segmentation of 3D volumes (CT, MR) and video sequences (ultrasound, endoscopy) is critical for medical tasks such as tumor quantification, organ volumetry, and surgical planning. However, manual segmentation of 3D data remains extremely challenging. All difficulties seen in 2D medical segmentation including low contrast, heterogeneous texture, imaging noise, and ambiguous boundaries persist and amplify in 3D. Moreover, errors in individual slices accumulate across volumes, disrupting anatomical continuity and maintaining consistency over hundreds of slices demands reasoning over 3D context and inter-slice relationships. Manual voxel-level annotation further compounds these issues, being both time-consuming and expertise-intensive.

Efforts to adapt FMs for medical imaging have largely relied on large-scale supervision. SAM-Med3D \cite{wang2025sam} and MedSAM-2 \cite{ma2025medsam2} fine-tune SAM-like models on massive curated medical corpora to achieve strong supervised performance but depend heavily on manual labeling and remain constrained by scanner-, protocol-, and anatomy-specific biases. Large supervised models such as TotalSegmentator~\cite{wasserthal2023totalsegmentator}, SegVol~\cite{du2024segvol}, and VISTA3D~\cite{he2025vista3d} further illustrate this dependence, while ultrasound-centric models like MemSAM, EchoFM, and EchoFlow~\cite{deng2024memsam,kim2024echofm,reynaud2025echoflow} address temporal coherence but remain modality-specific.

To address these limitations, we propose SynthFM-3D, a data-centric framework for training FMs using fully synthetic 3D data. The contributions of this work are as follows:

\begin{itemize}[noitemsep, topsep=0pt, leftmargin=*]
    \item SynthFM-3D systematically generates synthetic volumetric data that emulate the complexities of medical imaging, including variations in shape, texture, noise, and contrast, as well as 3D-specific challenges such as structural continuity across slices, appearance and disappearance of organs, and contextual consistency throughout volumes. All components of SynthFM-3D are mathematically parameterized, enabling fully controlled, reproducible, and scalable synthesis of diverse imaging conditions. 
    \item By generating paired synthetic images and ground-truth masks simultaneously, the proposed approach eliminates the reliance on large-scale manual annotations and enables the modality-agnostic adaptation of promptable segmentation models, such as SAM~2, for both medical volume and video segmentation tasks.
    \item SynthFM-3D showed strong generalization across multiple real-world medical segmentation tasks, including MRI, CT, and ultrasound modalities, validating its potential as a scalable alternative to supervised FMs.
\end{itemize}

%%%%%%%%%%%%%%%%%%%%%%%%%%%%%%%%%%%%%%%%%%%%%%%%%%%%%%%%%%%
\section{Methodology}
\vspace{-1em}
\subsection{Data generation strategy}
\vspace{-0.5em}
SynthFM-3D extends our prior 2D synthetic framework~\cite{sengupta2025synthfm} to volumes, targeting the anatomical variability and appearance characteristics seen in medical imaging modalities. The data generation pipeline has two stages: (i) 3D label masks are synthesized by extrapolating 2D multiclass masks via stochastic morphological transformations; (ii) image volumes are rendered from these 3D masks using structure-specific textures, contrast modulation, and boundary-aware blending. Formally, we denote the process by the mapping \(\Phi:(M,\theta_{\text{struct}},\theta_{\text{app}})\mapsto(Y,X)\), where \(M\in\mathbb{N}^{H\times W}\) is the seed 2D mask, \(\theta_{\text{struct}}\) and \(\theta_{\text{app}}\) are structural and appearance parameter sets, \(Y\in\mathbb{N}^{H\times W\times D}\) is the 3D label volume, and \(X\in[0,255]^{H\times W\times D}\) is the corresponding image. These parameters induce a reproducible mapping that yields anatomically consistent yet statistically diverse 3D datasets.

\vspace{-1em}
\begin{figure}[!htbp]
    \centering
    % \fbox{\rule{0pt}{5in}\rule{\linewidth}{0pt}}
    \includegraphics[width=\linewidth]{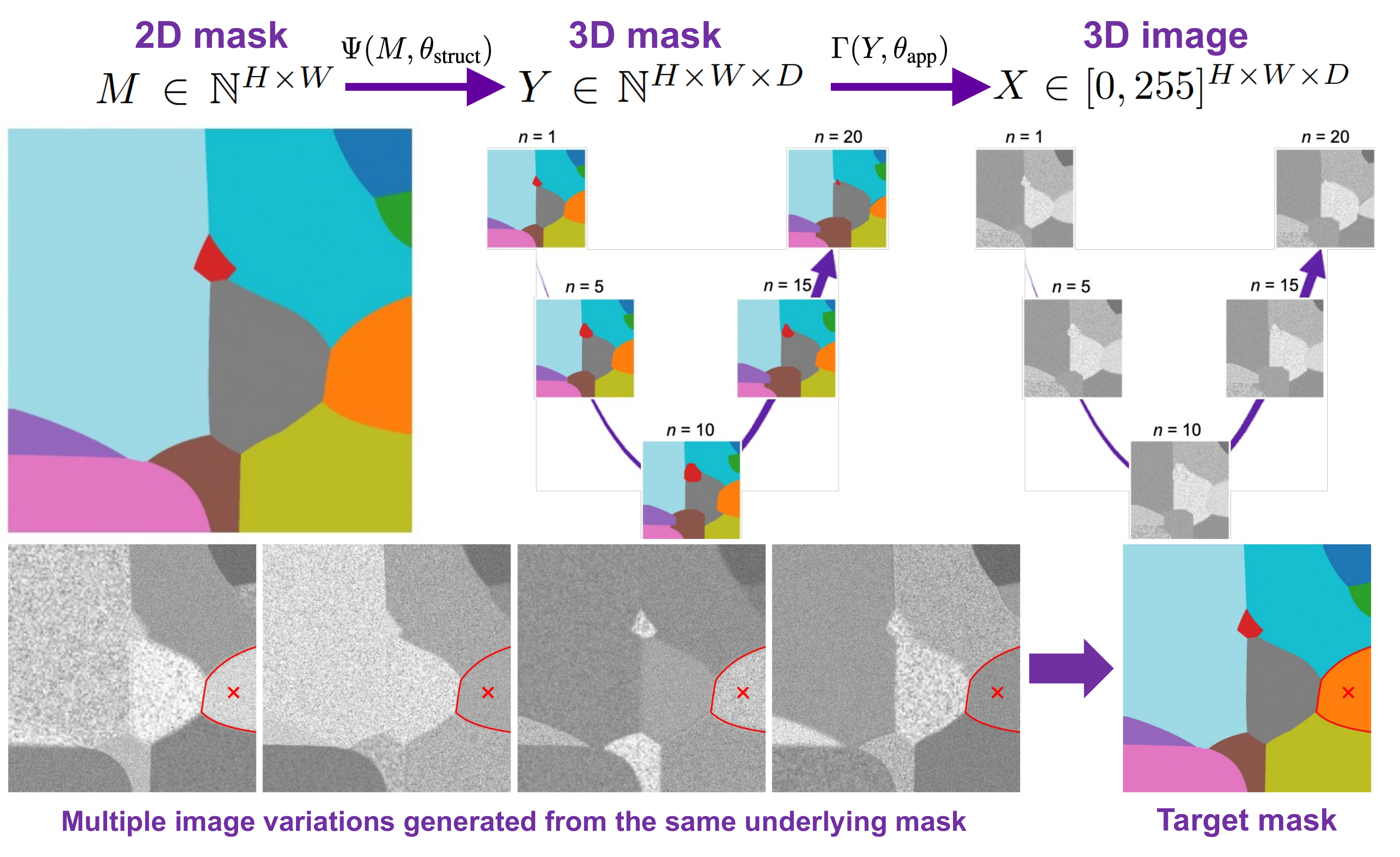} 
    \caption{
Overview of the SynthFM-3D data generation process. Top row illustrates the analytical mapping 
\(\Phi(M, \theta_{\text{struct}}, \theta_{\text{app}})\), where a 2D seed mask \(M\) is extrapolated into a 3D label volume 
\(Y = \Psi(M, \theta_{\text{struct}})\) and rendered into a corresponding image volume 
\(X = \Gamma(Y, \theta_{\text{app}})\). Bottom row shows the diversity from multiple texture and contrast 
variations generated from the same underlying mask, showing the controllable variability achieved through 
different parameters \((\theta_{\text{struct}}^{(i)}, \theta_{\text{app}}^{(j)})\).
}
\end{figure}

\vspace{-2.5em}
\subsection{Shape generation}
\vspace{-0.5em}
We start from a 2D multiclass segmentation mask \(M\in\mathbb{N}^{H\times W}\) produced via the boundary-aware module of SynthFM-2D~\cite{sengupta2025synthfm}, which uses SynthMorph~\cite{hoffmann2023synthmorph} to create label maps with \(10\)–\(15\) spatially coherent clusters. These clusters emulate adjacent anatomical regions with shared boundaries. Let \(L=\{1,\dots,K\}\) denote the foreground labels and \(B_\ell^{(1)}=\mathbf{1}[M=\ell]\) the initial binary mask for label \(\ell\).

We define a stochastic structural mapping \(\Psi:(M,\theta_{\text{struct}})\mapsto Y\) that evolves each \(B_\ell^{(1)}\) across depth \(d=1,\dots,D\) using iterative morphological dilation and erosion. For each label \(\ell\), we define structural parameters \(\theta_{\text{struct}}=\{g_\ell,s_\ell,d_\ell^\star,\pi_\ell\}\), where \(g_\ell\) and \(s_\ell\) are per-slice iteration counts for dilation and erosion, \(d_\ell^\star\) is the transition depth from dilation- to erosion-dominant evolution, and \(\pi_\ell\) indicates whether the structure evolves (\(\pi_\ell=1\)) or remains static (\(\pi_\ell=0\)). Recursively, for \(d\ge2\),
\vspace{-1.5em}
\begin{equation}
B_\ell^{(d)}=
\begin{cases}
\operatorname{dilate}\!\big(B_\ell^{(d-1)};K_{3\times3},g_\ell\big), & \pi_\ell=1,\ d<d_\ell^\star,\\[4pt]
\operatorname{erode}\!\big(B_\ell^{(d-1)};K_{3\times3},s_\ell\big), & \pi_\ell=1,\ d\ge d_\ell^\star,\\[4pt]
B_\ell^{(d-1)}, & \pi_\ell=0,
\end{cases}
\end{equation}
where \(K_{3\times3}\) is a \(3\times3\) structuring kernel. After all transformed binary masks \(\{B_\ell^{(d)}\}_{\ell=1}^{K}\) are generated, they are composited slice-wise into a label volume with fixed z-ordering by label index to ensure label exclusivity:
\vspace{-1em}
\begin{equation}
Y[:,:,d] = \sum_{\ell=1}^K \ell \cdot B_\ell^{(d)} \cdot \mathbf{1}[Y[:,:,d] = 0].
\end{equation}

This assembly guarantees non-overlap—each voxel belongs to a single class—and provides deterministic conflict resolution when structures intersect. The process allows anatomical structures to grow, shrink, persist, or appear and disappear across depth, producing heterogeneous yet spatially coherent 3D shapes that reflect the continuity and variability characteristic of real volumetric medical data.
\vspace{-1.5em}
\subsection{Contrast and texture generation}
\vspace{-0.5em}
After constructing the 3D label volume \(Y\in\mathbb{N}^{H\times W\times D}\), we synthesize \(X\in[0,255]^{H\times W\times D}\) by assigning each label a texture field and composing slices with background variation and boundary smoothing, parameterized by \(\theta_{\text{app}}\). For each \(\ell\in L\), we generate a texture,
\vspace{-0.5em}
\begin{equation}
T_\ell=\frac{1}{Z_\ell}\sum_{i=1}^{n_\ell}\alpha_i\,\mathcal{N}_i(0,0.01),\quad n_\ell\sim\mathrm{Uniform}(1,4),
\end{equation}

where \(n_\ell \) defines the number of scales, \(\{\alpha_i\}\) are random mixture weights normalized by \(Z_\ell=\sum_i\alpha_i\), and each \(\mathcal{N}_i\) is a Gaussian noise layer spatially rescaled by a zoom-factor \(z_i=2^{n_\ell}\) to capture multi-scale texture. Each texture is scaled by a random intensity coefficient \(I_\ell\sim\mathrm{Uniform}(0.1,0.9)\), reflecting plausible tissue brightness variations.

For each slice \(d = 1, \dots, D\), a background field 
\(B_d \sim \mathcal{N}(0.5, 0.1)\) is perturbed with subtle random variation 
\(\epsilon_d \sim \mathcal{N}(0, 0.02)\) to introduce spatial heterogeneity:
\vspace{-0.5em}
\begin{equation}
\tilde{B}_d = \mathrm{clip}(B_d + \epsilon_d, 0, 1).
\end{equation}
Each structure’s binary mask \(B_\ell^{(d)} = \mathbf{1}[Y[:,:,d] = \ell]\) 
is softened by Gaussian blurring to emulate partial-volume effects:
\vspace{-0.5em}
\begin{equation}
\tilde{B}_\ell^{(d)} = G_{\sigma_\ell}(B_\ell^{(d)}),
\quad \text{where} \quad \sigma_\ell \sim \mathrm{Uniform}(2, 8).
\end{equation}

The slice intensity field is then composed as:
\vspace{-0.5em}
\begin{equation}
X[:,:,d] = \tilde{B}_d + \sum_{\ell=1}^{K} I_\ell \, T_\ell \, \tilde{B}_\ell^{(d)},
\end{equation}
where texture, intensity, and boundary blur collectively define the appearance of each structure.
Finally, the complete 3D image volume is normalized to the 8-bit range:
\begin{equation}
X = 255 \cdot \frac{X - \min(X)}{\max(X) - \min(X)}.
\end{equation}

This formulation ensures realistic contrast, smooth transitions, and background variation across slices,
yielding anatomically coherent and artifact-free synthetic volumes suitable for training and evaluation 
of segmentation models across multiple imaging modalities.

\vspace{-1.5em}
\subsection{Controlling variation in synthetic data}
\vspace{-0.5em}
SynthFM-3D enables systematic control over both structural and appearance diversity, allowing multiple volumetric realizations to be derived from a single 2D seed mask \(M\). Diversity is controlled at two levels. Structural variation samples independent \(\theta_{\text{struct}}^{(i)}\sim\mathcal{P}_{\text{struct}}\) (per-slice dilation/erosion iteration counts and transition depths \(d_\ell^\star\)) to produce distinct yet coherent shapes \(Y^{(i)}=\Psi(M,\theta_{\text{struct}}^{(i)})\). Appearance variation samples \(\theta_{\text{app}}^{(j)}\sim\mathcal{P}_{\text{app}}\) (texture mixtures, intensity scales, blur strengths, background statistics) to render multiple images for a fixed mask: \(X^{(j)}=\Gamma\!\big(Y,\theta_{\text{app}}^{(j)}\big)\). Together,
\vspace{-0.5em}
\begin{equation}
\big(X^{(i,j)},Y^{(i)}\big)=\Phi\!\big(M,\theta_{\text{struct}}^{(i)},\theta_{\text{app}}^{(j)}\big),
\end{equation}
yielding statistically independent samples that preserve anatomical plausibility while enabling broad coverage of structural and appearance variability. Unlike traditional augmentation, which perturbs existing data, this framework synthesizes new, statistically independent samples that enhance model robustness to anatomical and modality variations.
\vspace{-1.5em}
\subsection{Model architecture and Training details}
\vspace{-0.5em}
The SAM~2.1 (Hiera-B+) checkpoint was fine-tuned on 10{,}000 synthetic 3D cases generated using SynthFM-3D for 40 epochs, following all default hyperparameters. This configuration was designed for a fair comparison with the original pre-trained SAM 2 and to ensure that the SynthFM-3D–trained checkpoints function as drop-in replacements for standard SAM 2 models. During evaluation, two click–prompt strategies were employed: (i) a single +ve click (placed at the centroid of the target object) and (ii) three +ve clicks (one centroid and two additional clicks distributed randomly within the target region). All clicks were placed on the starting frame to initialize the segmentation and then propagated through subsequent slices of the sequence or volume.

\begin{figure*}[!htbp]
    \centering
    \includegraphics[width=\textwidth]{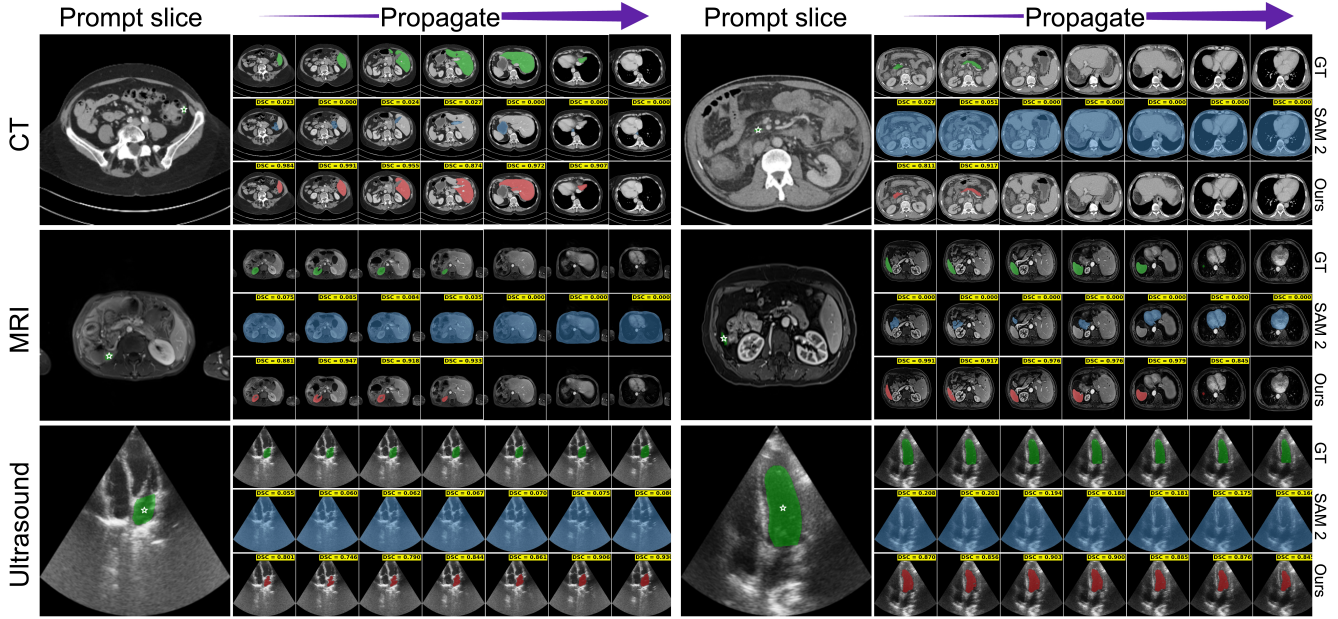} 
    \caption{Qualitative comparison of segmentation propagation between SAM 2 and SynthFM-3D. Each row corresponds to a distinct modality and two examples per modality are shown. For each case, the first column depicts the \textit{prompt slice}, where the positive prompt location is marked by a green star (\(\color{green}{\star}\)); subsequent columns show segmentation \textit{propagation} across slices.}
    \vspace{-1em}
\end{figure*}

%%%%%%%%%%%%%%%%%%%%%%%%%%%%%%%%%%%%%%%%%%%%%%%%%%%%%%%%%%%
\vspace{-2em}
\section{Experiments and Results}
\vspace{-1em}
\subsection{Datasets and Evaluation Metrics}
\vspace{-0.5em}
SynthFM-3D was evaluated on the following clinical datasets spanning multiple medical imaging modalities: TotalSegmentator~\cite{ref:total,totalsegmentatormri} and AMOS~\cite{ref:amos} for CT/MR, and CAMUS~\cite{ref:camus} for echocardiography ultrasound. CT/MR evaluations included abdominal structures viz. aorta, gallbladder, kidney, liver, pancreas, prostate, and spleen. Ultrasound evaluations targeted cardiac structures viz. left atrium and left ventricle endometrium. 
Dice similarity coefficient (DSC) was used as the primary metric. Statistical significance was assessed using a Student’s paired \(t\)-test between methods.

\subsection{Zero-shot Comparison with SAM~2}
\vspace{-0.5em}
Zero-shot evaluation was performed to assess the ability of SynthFM-3D to bridge the domain gap between natural and medical image segmentation. Pre-trained SAM~2.1, having never been exposed to medical data, serves as the out-of-domain baseline. By fine-tuning SAM~2.1 (Hiera-B+) exclusively on synthetic 3D data generated using SynthFM-3D and evaluating it on real medical data without any further adaptation, we directly quantify the effect of synthetic data on cross-domain generalization. 
This setup provides a controlled comparison between SAM~2.1 with and without synthetic training, isolating the contribution of synthetic data toward improving segmentation performance on CT, MR, ultrasound volumes.
\vspace{-0.5em}

\input{tables/zeroshot}
\vspace{-1em}
From Table~\ref{tab:zeroshot}, we see that across all datasets and modalities, SynthFM-3D demonstrates clear and consistent gains over the pre-trained SAM~2.1 baseline under both \((1,0)\) and \((3,0)\) prompt configurations. For CT, DSC improvements of $20$-$45$ points are observed across most organs, with the largest gains for the pancreas ($+23$), prostate ($+38$), and spleen ($+39$), indicating enhanced adaptation to structural and contrast heterogeneity. 
In MR, improvements of $12$-$45$ DSC are observed across nearly all organs, particularly for the bladder ($+65$) and spleen ($+40$), maintaining robustness to modality-specific intensity distributions. The largest performance jumps occur in ultrasound, with $+38$ and $+36$ DSC improvements for the LA and LV endocardium, respectively, reflecting SynthFM-3D’s capacity to generalize to unseen low-contrast modalities. Nearly all gains are statistically significant, confirming that the improvements are systematic and not due to random variation.

\vspace{-1em}
\subsection{Comparison with Supervised SOTA (SAM-Med3D)}
\vspace{-0.5em}
To evaluate how closely synthetic pretraining can approach fully supervised performance, we benchmark SynthFM-3D against the supervised foundation model SAM-Med3D. SAM-Med3D, trained on diverse modalities but excluding ultrasound, shares the same click-prompt interface as SynthFM-3D, making it an appropriate baseline. Both models were evaluated in a strict zero-shot setting on the unseen CAMUS dataset using \((1,0)\) prompts. SynthFM-3D clearly outperforms SAM-Med3D on both cardiac structures (Table~\ref{tab:sammed3d}). DSC improves by 2–3×, with the largest gains for the LV Endo. 
% Qualitative results (\hl{figure}) show that SynthFM-3D produces smoother, anatomically realistic segmentations across slices, while SAM-Med3D often under-segments or misses structures.
\vspace{-0.5em}
\input{tables/sammed3d}
\vspace{-2em}
\subsection{Few-Shot Structure-specific Fine-tuning}
\vspace{-0.5em}
A few-shot fine-tuning study was performed to assess performance when limited annotations are available. For this, SynthFM-3D was compared to SAM 2, SAM-Med3D, and additionally nnU-Net~\cite{isensee2021nnu}. 
All models were fine-tuned on the CAMUS dataset using \((1,0)\) prompts (except nnU-Net) using $2$ and $5$ annotated cases for each cardiac structure and evaluated on the same structure. Results (Table~\ref{tab:fewshot}) show that SynthFM-3D consistently achieves the highest Dice scores among promptable models, outperforming SAM~2 and SAM-Med3D by large margins in the extreme few-shot regimes (\(+20\)–\(+25\)\% DSC with only 2–5 annotations). These findings demonstrate that synthetic pretraining substantially enhances label efficiency and structural delineation under limited supervision. Notably, for LV-endo, SynthFM-3D surpasses nnU-Net for 2 cases and approaches its performance for 5 cases. These results highlight the strong inductive bias transferred from synthetic pretraining and its effectiveness in data-scarce clinical settings.
\vspace{-1em}
\input{tables/fewshot}
\vspace{-3em}
\section{Conclusion}\label{sec:conclusion}
\vspace{-1em}
This work introduced SynthFM-3D, a systematic framework that captures the underlying statistical and structural properties of medical imaging by analytically modeling variability in anatomy, boundary definition, contrast, and noise to enable training of promptable FMs without real annotated data. Unlike prior adaptations that rely on large curated medical corpora, SynthFM-3D adopts a \emph{foundational data} approach, positioning data synthesis as the core enabler of generalization rather than extensive human supervision. Zero-shot evaluations and comparisons with supervised SOTA across different imaging modalities showed that training FMs using SynthFM-3D matched or exceeded supervised medical counterparts, underscoring the effectiveness of analytically generated synthetic data as a scalable, annotation-free training strategy across medical imaging modalities.

\bibliographystyle{IEEEbib}
\bibliography{main}

\end{document}

%% file: tables/zeroshot.tex
\begin{table}[!htbp]
\centering
\scriptsize
\setlength{\tabcolsep}{3pt}
\renewcommand{\arraystretch}{1}
\caption{Quantitative results (DSC) for zero-shot segmentation across modalities and anatomical structures under (1,0) and (3,0) prompts. Significance is shown as \textbf{\underline{xx.xx}} for $p<0.001$, \textbf{xx.xx} for $0.001<p<0.05$, and xx.xx for $p>0.05$.}
\vspace{-1em}
\label{tab:zeroshot} 
\begin{tabular}{c|c|cc|cc} % <-- no '|' at the extremes
% \hline
\toprule
\multirow{2}{*}{\textbf{Modality}} &
\multirow{2}{*}{\textbf{Structure}} &
\multicolumn{2}{c|}{\textbf{(1,0)}} &
\multicolumn{2}{c}{\textbf{(3,0)}} \\
\cline{3-6}
& & \textbf{SAM 2} & \textbf{SynthFM-3D} & \textbf{SAM 2} & \textbf{SynthFM-3D} \\
\hline
\multirow{9}{*}{CT}
 & Aorta           & 68.18 & \textbf{72.87}             & 71.67 & 74.49 \\
 & Bladder         & 23.44 & \textbf{\underline{66.71}} & 28.34 & \textbf{\underline{66.89}} \\
 & Gallbladder     & 31.09 & \textbf{\underline{66.31}} & 38.98 & \textbf{\underline{69.25}} \\
 & Kidney (L)      & 63.54 & \textbf{\underline{79.95}} & 71.88 & \textbf{\underline{84.30}} \\
 & Kidney (R)      & 58.61 & \textbf{\underline{79.35}} & 69.60 & \textbf{\underline{81.92}} \\
 & Liver           & 48.33 & \textbf{\underline{73.49}} & 56.93 & \textbf{\underline{77.02}} \\
 & Pancreas        & 19.52 & \textbf{\underline{42.44}} & 22.09 & \textbf{\underline{43.68}} \\
 & Prostate        & 14.78 & \textbf{\underline{52.26}} & 23.26 & \textbf{\underline{54.31}} \\
 & Spleen          & 44.74 & \textbf{\underline{83.24}} & 58.60 & \textbf{\underline{84.80}} \\
\hline
\multirow{9}{*}{MR}
 & Aorta           & 44.25 & \textbf{\underline{60.91}} & 47.40 & \textbf{\underline{60.29}} \\
 & Bladder         &  7.21 & 72.18                      & 11.70 & 76.33 \\
 & Gallbladder     & 22.69 & \textbf{\underline{48.12}} & 32.23 & \textbf{\underline{49.47}} \\
 & Kidney (L)      & 52.50 & \textbf{63.95}             & 59.96 & 63.75 \\
 & Kidney (R)      & 56.83 & \textbf{\underline{69.80}} & 63.24 & 69.57 \\
 & Liver           & 42.58 & \textbf{\underline{61.76}} & 48.26 & \textbf{\underline{60.57}} \\
 & Pancreas        &  9.37 & \textbf{\underline{32.54}} & 12.97 & \textbf{\underline{31.56}} \\
 & Prostate        & 26.18 & \textbf{40.15}             & 30.50 & \textbf{45.96} \\
 & Spleen          & 31.80 & \textbf{\underline{71.90}} & 43.16 & \textbf{\underline{72.97}} \\
\hline
\multirow{2}{*}{US}
 & LA              & 19.49 & \textbf{\underline{57.35}} & 21.40 & \textbf{\underline{54.68}} \\
 & LV Endocardium  & 27.47 & \textbf{\underline{63.80}} & 28.63 & \textbf{\underline{60.90}} \\
% \hline
\bottomrule
\end{tabular}
\end{table}

%% file: tables/sammed3d.tex
\begin{table}[!htbp]
\centering
\scriptsize
\setlength{\tabcolsep}{3pt}
\renewcommand{\arraystretch}{1}
\caption{Quantitative results (DSC) for comparison with supervised SOTA. Significance is shown as \textbf{\underline{xx.xx}} for $p<0.001$, \textbf{xx.xx} for $0.001<p<0.05$, and xx.xx for $p>0.05$.}
\vspace{-0.5em}
\label{tab:sammed3d} 
\begin{tabular}{c|c|c|cc} % internal verticals only
% \hline
\toprule
\multirow{2}{*}{\textbf{Modality}} &
\multirow{2}{*}{\textbf{Dataset}} &
\multirow{2}{*}{\textbf{Structure}} &
\multicolumn{2}{c}{\textbf{(1,0)}} \\
\cline{4-5}
& & & \textbf{SAM-Med3D} & \textbf{SynthFM-3D} \\
\hline
\multirow{2}{*}{US}
 & \multirow{2}{*}{CAMUS}
 & LA        & 25.81 & \textbf{\underline{57.35}} \\
 &            & LV Endocardium & 21.60 & \textbf{\underline{63.80}} \\
 % &            & LV Epicardium  &  1.34 & \textbf{\underline{21.67}} \\
% \hline
\bottomrule
\end{tabular}
\end{table}

%% file: tables/fewshot.tex
% \begin{table}[!htbp]
% \centering
% \scriptsize
% \setlength{\tabcolsep}{4pt}
% \renewcommand{\arraystretch}{1.05}
% \caption{CAMUS results under (1,0) prompts. Values are DSC\,(\%). The highest value in each row is highlighted in bold.}
% \begin{tabular}{c|c|ccc|c}
% \toprule
% \textbf{Object} & \textbf{Cases} & \textbf{SAM2.1} & \textbf{nnU-Net} & \textbf{SAM-Med3D} & \textbf{SynthFM-3D} \\
% \midrule
% \multirow{4}{*}{LA} 
%  & 2  & 41.56 & 40.04 & 44.04 & \textbf{64.72} \\
%  & 5  & 38.68 & 46.39 & 44.24 & \textbf{64.97} \\
%  %& 10 & \textbf{78.76} & 57.99 & 44.42 & 70.10 \\
%  %& 20 & 69.32 & 69.75 & 44.48 & \textbf{72.90} \\
% [2pt]
% \midrule
% \multirow{4}{*}{LV Endocardium} 
%  & 2  & 58.31 & 51.95 & 41.93 & \textbf{70.39} \\
%  & 5  & 55.12 & \textbf{74.73} & 41.82 & 72.41 \\
%  %& 10 & 70.98 & 71.30 & 41.79 & \textbf{74.65} \\
%  %& 20 & 77.04 & \textbf{81.66} & 41.66 & 75.92 \\
% %\bottomrule
% \label{tab:fewshot}
% \end{tabular}
% \end{table}

\begin{table}[!htbp]
\centering
\scriptsize
\setlength{\tabcolsep}{3pt}
\renewcommand{\arraystretch}{1}
\caption{Quantitative results (DSC) for few-shot fine-tuning. Best results are in \textbf{bold}.}
\vspace{-0.5em}
\label{tab:fewshot}
\begin{tabular}{c|c|cccc} 
\toprule
\multirow{2}{*}{\textbf{Structure}} &
\multirow{2}{*}{\textbf{Cases}} &
\multicolumn{4}{c}{\textbf{Model}} \\
\cline{3-6}
& & \textbf{SAM2.1} & \textbf{SAM-Med3D} & \textbf{nnU-Net} & \textbf{SynthFM-3D} \\
\hline
\multirow{2}{*}{LA} 
 & 2  & 41.56 & 44.04 & 40.04 & \textbf{64.72} \\
 & 5  & 38.68 & 44.24 & 46.39 & \textbf{64.97} \\
\hline
\multirow{2}{*}{LV Endocardium} 
 & 2  & 58.31 & 41.93 & 51.95 & \textbf{70.39} \\
 & 5  & 55.12 & 41.82 & \textbf{74.73} & 72.41 \\
\bottomrule
\end{tabular}
\end{table}